\documentclass[a4paper, oneside, twocolumn, notitlepage, 10pt]{extarticle_ecoc}
\usepackage{ecoc}
\usepackage{subcaption}

\usepackage{etoolbox}
\AtBeginDocument{%
  \providebibmacro*{cite:init}{}%
  \providebibmacro*{cite:comp}{\usebibmacro{cite}}%
  \providebibmacro*{cite:dump}{}%
}

\begin{document}
\selectlanguage{English}    
\title{Precise Localization of High-Voltage Breakdown Events using $\phi$-Optical Time-Domain Reflectometry on an Optical Ground Wire}
\author{
    Konstantinos Alexoudis\textsuperscript{(1,2)}, Luke Silvestre\textsuperscript{(3)}, Tom Huiskamp\textsuperscript{(3)}, Jasper Müller\textsuperscript{(1)}, Vincent Sleiffer\textsuperscript{(1)}\\
    Florian Azendorf\textsuperscript{(1)}, Sander Jansen\textsuperscript{(1)}, Chigo Okonkwo\textsuperscript{(2)} and Tom Bradley\textsuperscript{(2)}
}

\maketitle                  

\begin{strip}
    \begin{author_descr}

        \textsuperscript{(1)} Adtran, Frauenhoferstraße 9a, 82152 Planegg, Germany,
        \textcolor{blue}{\uline{konstantinos.alexoudis@adtran.com}}
        
        \textsuperscript{(2)} High-Capacity Optical Transmission Laboratory, Eindhoven University of Technology, Netherlands
        
        \textsuperscript{(3)} Electrical Energy Systems Research Group, Eindhoven University of Technology, Netherlands

    \end{author_descr}
\end{strip}

\renewcommand\footnotemark{}
\renewcommand\footnoterule{}


\begin{strip}
    \begin{ecoc_abstract}        
        We present $\phi$-OTDR for detecting and localising full spark-gap breakdowns by analysing backscattered light phase and frequency-domain signatures during high-voltage discharges synchronised with oscilloscope-recorded events. Measuring sub-kHz confirms clear discharge signatures and acoustic reconstruction over long links with $\approx 10$~m spatial resolution. \textcopyright2025 The author(s)
    \end{ecoc_abstract}
\end{strip}

\section{Introduction}
Optical fibres have been extensively demonstrated for high-capacity optical communication networks. Serving this purpose for (critical) communication or possibly commercial activities, they have been deployed in power utility networks and typically integrated into or wrapped around ground wires between electricity pylons   ~\cite{OPGW:05, OPGW:24}. Recently, monitoring the security and reliability of critical infrastructure has been an increasing priority, focused on lowering maintenance costs and utilising the power network at maximum capacity.
Optical ground wires (OPGW) may be ideally suited for this ~\cite{Verizon:24, NSN:18, Mazur:23, Discharge:19}.

High-voltage (HV) insulation faults, ranging from partial discharge (PD) to full spark-gap breakdowns, are the earliest indicators of component failures such as cables, switchgear, transformers, and overhead lines, quickly leading to equipment damage. This damage results in outages, leading to high repair costs and safety hazards~\cite{Discharge:19, Discharge:Gen}. Early detection of these discharges can establish targeted proactive maintenance activities to avoid catastrophic damage, allowing longer asset life and lower cost of operation.

Traditional techniques, such as insulation resistance testing and cable fault location techniques, are often not immune to electromagnetic interference (EMI) and are only capable of monitoring single points on cables or overhead lines. For comparison, distributed sensing, for example using phase-sensitive optical time-domain reflectometry ($\phi$-OTDR) allows distributed sensing along an optical fibre and thus can be used for sensing between pylons with a single interrogator unit ~\cite{Adtran:25}.

Recent work has shown that $\phi$-OTDR achieves metre-scale detection of partial discharges and sub-metre localisation of switchgear breakdowns using custom 3D printed sensing elements \cite{kirkcaldy:2021, chen:2020}.

In this paper, we built on these findings by investigating the feasibility of $\phi$-OTDR to detect and localise complete spark-gap breakdown on long-distance field-deployed OPGW under realistic noise conditions. Therefore, we quantify the sensitivity of different fibre configurations and synchronise OTDR measurements with oscilloscope recorded HV-triggers for microsecond accurate time localisation. We tested different pulse repetition rates while adding additional Gaussian noise to demonstrate that low repetition rates suffice for the detection of electric discharge, confirming the feasibility of $\phi$-OTDR for monitoring the electric grid over long distances (50-100~km). 

\section{Experimental Setup}
\begin{figure*}[t!]
    \centering    \includegraphics[width=\linewidth]{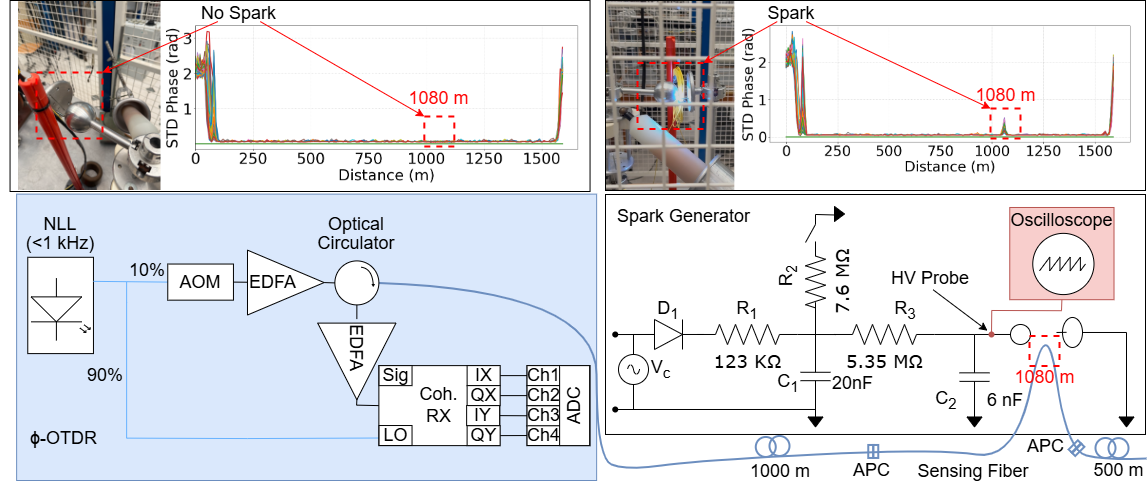}
    \caption{Experimental setup: schematic of the $\phi$-OTDR interrogator (left), fingerprint showing phase difference along the 1.5 km test fibre with the arc location, and schematic of the testbed as well as a picture of the spark-gap with sensing fibre (right).}
    \label{fig:setup}
\end{figure*}

To emulate discharge events and measure them, the setup in Fig. \ref{fig:setup} was built. A $\phi$-OTDR interrogator was located outside a grounded metal cage to reduce EMI. This interrogator is built using a <1 kHz narrow-linewidth laser (NLL), which goes through a 90/10~split. The 90\%  branch is intensity modulated using an acousto-optic modulator (AOM) that we used to generate pulses with repetition rates of 400~Hz to 15~kHz and pulse widths of 50 and 100~ns and a gauge length of 10 metres. Afterwards, the signal is amplified using an erbium-doped fibre amplifier (EDFA). The amplified light, with +10 dBm peak power, is then sent through an optical circulator connected with angled physical connectors (APC), into a 1~km launch fibre, which passes through the cage mesh into the enclosure. Inside, this fibre is connected to the sensor fibre positioned next to the spark gap and then leads into the 500 m of tail fibre. The Rayleigh backscattered signal is first amplified using an EDFA, before being fed to a standard coherent receiver setup, which uses the 10\% branch from the NLL as the local oscillator (LO). After optical-to-electrical conversion via photodetectors, the resulting analogue voltages are sampled using analogue-to-digital converters (ADC) with a sampling rate of 250 MS/s. 
For each measurement, we recorded $\phi$-OTDR traces for 5~s to establish a baseline, then initiated the spark generation set-up and captured 10~s of data, followed by 5~s of post-event logging. During the same time, a Rigol DS1204B oscilloscope logged the HV probe waveform.

The spark generation set-up uses step-up transformers operating at 50~Hz to produce an HV charge denoted $V_{c}$, see Fig.~\ref{fig:setup}. The HV diode $D_{1}$ with resistor $R_{1}$ and capacitor $C_{1}$ behaves as a half-wave rectifier with a sufficiently low voltage ripple for HV charging. The voltage produced by the half-wave rectifier is used to charge the discharge capacitor $C_{2}$ through the charging resistor $R_{3}$. The spark generation set-up had an average time delay between breakdown events of approximately 0.114 $s$, making it easy to distinguish between individual breakdown events. The spark gap where the discharge occurs is a sphere and needle geometry with a screw used as the needle. The discharges are recorded using a Rigol DS1204B oscilloscope measuring the charge and discharge of capacitance $C_{2}$, see Fig.~\ref{fig:setup}. The HV probe used for measurements is the PVM-5 North Star probe, which is a combination of resistive and capacitive voltage divider probes. After obtaining the required measurements, the spark generation set-up is turned off and the switch connected to the grounding resistor $R_{2}$ is closed to safely dissipate any remaining charge in $C_{1}$ and $C_{2}$.

We evaluated two different tight-buffered cables as sensor fibre, a laboratory-grade, single-core 3~mm-jacketed SSMF and an industry-standard OPGW cable built around an aramid strength member with four tight-buffered fibres helically wrapped and over-sheathed in a robust plastic jacket. This cable is deployed in a power utility network in Europe, wrapped around the ground wire. Each cable was first laid straight to record its baseline backscatter profile and then coiled next to the discharge location, with 13 turns for the flexible 3 mm SSMF and 4 turns for the stiffer OPGW cable, to evaluate the impact of bending on discharge-detection sensitivity. 

\section{Results and discussion}
We evaluated the four different measurement configurations: lab fibre, straight and coiled (13 turns), and OPGW, straight and coiled (4 turns). 
All configurations produced a visible increase in the measured phase change at the discharge location, as illustrated in the inset of Fig.~\ref{fig:setup} with the phase standard deviation as a function of time. The largest response was observed for the coiled lab fibre, followed by the straight lab fibre, with the coiled and straight OPGW yielding smaller but detectable phase changes of up to 0.5 radians. The lab fibre exhibits better mechanical coupling to the pressure wave generated by electrical discharge than a typical OPGW, which may be attributed to the helical arrangement of its internal fibres. Given that it is typically deployed in production networks, the straight OPGW was selected for further analysis. These results were obtained using a pulse repetition frequency of 15~kHz.

\begin{figure}[t!]
    \centering
    \includegraphics[width=\linewidth]{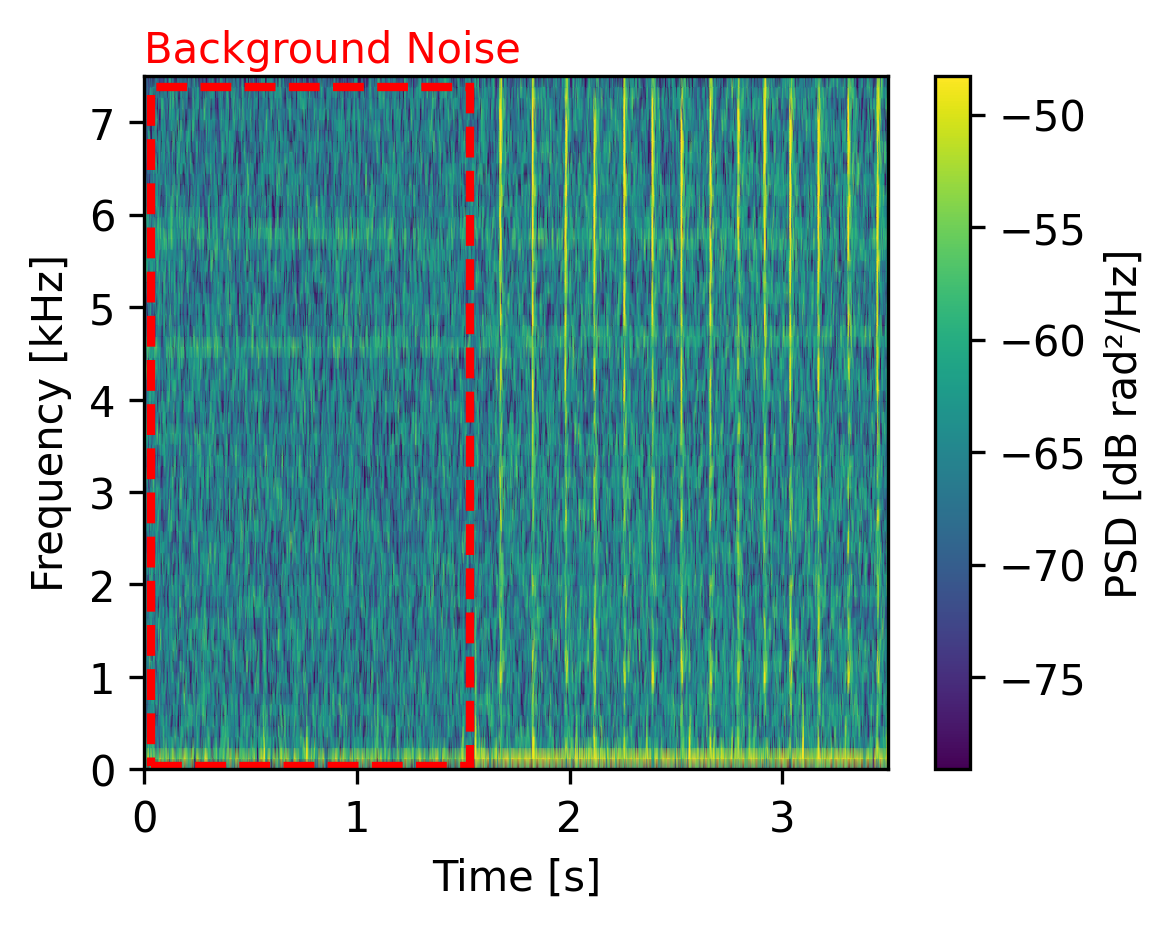}
    \caption{Time–frequency spectrogram at the spark location, with discharge: sharp vertical bands indicate breakdown events. The persistent horizontal bands at around 4.5 and 5.5~kHz are induced by laboratory background noise and present in reference measurements as shown in the red box.}
    \label{fig:spectrogram}
\end{figure}

Fig.~\ref{fig:spectrogram} shows the time-frequency spectrogram recorded by the $\phi$-OTDR at the known discharge location, approximately 1080~m from the interrogator, with an uncertainty of +/- 5~m. It shows the breakdown events as distinct vertical bands, with spectral energy extending from a few hertz up to the 7.5~kHz Nyquist limit. In addition to these vertical lines, two persistent horizontal lines appear around 4.5 and 5.5~kHz. Furthermore, elevated background noise appears across the lower frequency range as horizontal bands around 100 Hz, originating from the doubling effect of the 50~Hz power supply during each discharge and near 200~Hz, corresponding to the speed of our device fan. The vertical event bands broaden noticeably at lower frequencies, reflecting the longer duration and lower attenuation of low-frequency acoustic components. 
The persistence of the horizontal lines already present before the discharge events, indicated by the red box, confirms that they originate from background noise. Our measurement results without discharge events also support this. 

\begin{figure}[b!]
    \vspace{-1.8\baselineskip}
    \centering
    \includegraphics[width=\linewidth]{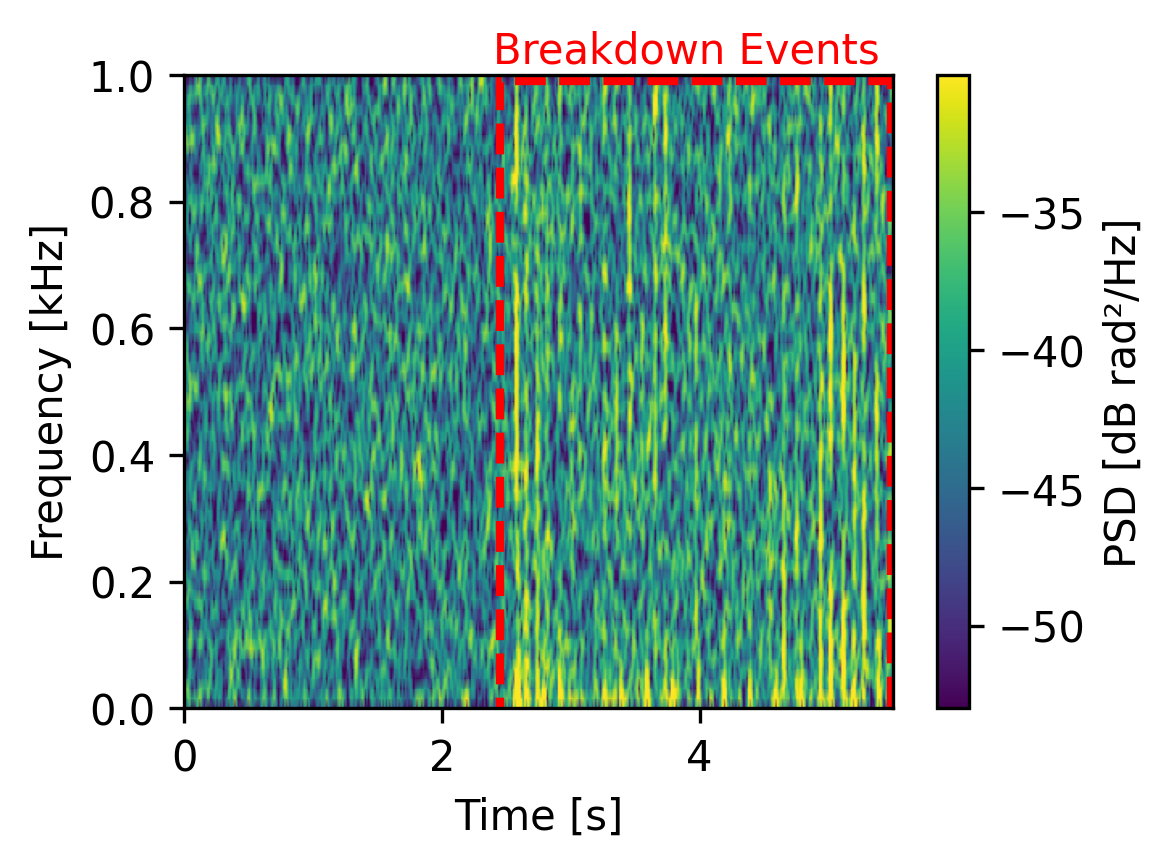}
    \caption{Time–frequency spectrogram at a 2~kHz pulse rate with an additional 5.4~dB AWGN across the 50 Hz to 1~kHz band. Despite the elevated low-frequency noise floor, the discharge-induced vertical bands remain visible indicating capabilities of long range detection under field-like conditions.}
    \label{fig:spectrogram_2000hz}
\end{figure}

We also verified this by reconstructing the audio from the measurements, where the sound of discharge events, as well as the background noise of the laboratory, could be identified. At a pulse repetition frequency of 1~kHz, the sparks are faintly audible but higher-frequency components are missing. Increasing the rate to 2~kHz produces a sharper, more defined acoustic signature, and at 5~kHz the reconstructed audio reveals post-discharge ringing at each event peak. 

Given that higher pulse repetition rates improve temporal resolution and capture higher-frequency content but reduce the maximum sensing range, we also tested sub-10~kHz pulse rates to verify that breakdown events remain detectable over 10 to 100 kilometre distances as required in actual power utility networks.

To assess applicability for field deployments, we elevated the noise floor of a measurement with a pulse repetition rate of 2~kHz (corresponding to a 50~km measurable reach), by adding additional white Gaussian noise (AWGN) of 5.4~dB. The sparks could still be detected in the reconstructed audio, as well as seen in the spectrogram (see Fig. \ref{fig:spectrogram_2000hz}). Audible, the sparks were still noticeable when adding up to 9~dB AWGN, although the spectogram blurs. Dedicated field trials are required to validate this performance under actual environmental and network noise conditions.

\begin{figure}[!tpb]
    \centering
    \includegraphics[width=\linewidth]{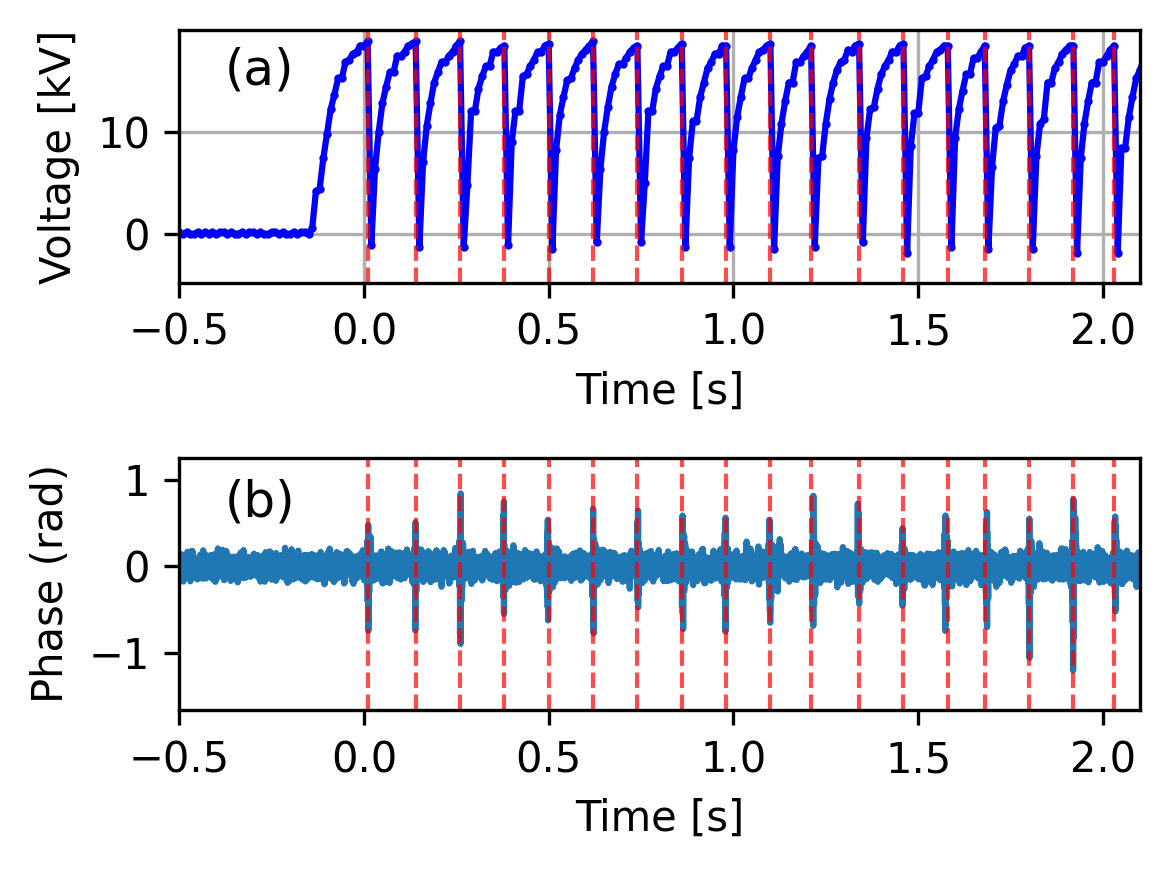}
    \caption{HV-probe oscilloscope pulses for multiple consecutive discharges (a) and overlay of $\phi$-OTDR event times (b). The red dashed lines correspond to temporal matches in the electrical breakdown and phase spikes detected via the OTDR. }
    \label{fig:otd_scope_sync}
\end{figure}

Finally, to validate the timing accuracy of our measurements, we synchronised the measured peaks with the breakdowns recorded on a Rigol DS1204B oscilloscope. The discharge times occurring at an average rate of 8.74~Hz were identified in the oscilloscope measurement shown in Fig.~\ref{fig:otd_scope_sync}~(a), marked by dashed lines. These discharge times are shown in the time-synchronised $\phi$-OTDR measurement Fig.~\ref{fig:otd_scope_sync}~(b), the discharge times align with the large phase change oscillations observed in the $\phi$-OTDR measurement.

\section{Conclusion}
We have demonstrated that a $\phi$-OTDR interrogator reliably detects and localises full spark electrical breakdowns on both a laboratory-grade SSMF and a field-deployed OPGW with a spatial resolution of $\approx 10$~m. Phase changes of up to 2~radians for the coiled lab fibre and 0.5~radians for the straight OPGW are observed. Uncoiled OPGW produces a clear phase-difference peak at the discharge location and distinct vertical bands in the frequency domain. Coiling or using lab fibre enhances the sensitivity. Synchronisation with oscilloscope data shows microsecond accuracy between spikes measured by $\phi$-OTDR and actual electrical breakdowns. These findings indicate that OPGW can serve not only as communications links but also as distributed EMI-resilient sensors for predictive fault detection in power networks.

\clearpage
\section{Acknowledgments}
This work has been partially funded by the German Federal Ministry of Education and Research in the project HYPERCORE (\#16KIS2098). We also acknowledge the Bilateral Project "DistraSignalSense" between the Eindhoven University of Technology, The Netherlands, and Adtran Networks SE.

\defbibnote{myprenote}

\printbibliography[prenote=myprenote]

\vspace{-4mm}

\end{document}